%% file: main.tex
\documentclass[10pt]{article}

\usepackage[margin=1in]{geometry}
\usepackage[T1]{fontenc}
\usepackage[utf8]{inputenc}
\usepackage{amsmath}
\usepackage{amssymb}
\usepackage{booktabs}
\usepackage{enumitem}
\usepackage{float}
\usepackage{microtype}
\usepackage{tabularx}
\usepackage[round]{natbib}
\usepackage{xcolor}
\usepackage{hyperref}

\hypersetup{
  colorlinks=true,
  linkcolor=blue!55!black,
  citecolor=blue!55!black,
  urlcolor=blue!55!black,
  pdftitle={Authority at Commit Time: Reject-and-Rerun Semantics for Governed Agentic Systems},
  pdfauthor={Jesus Salas (jesus.salas@gmail.com), Independent Researcher},
  pdfsubject={Authoritative admission of concurrent agent proposals under changing institutional state}
}

\newcommand{\current}{\ensuremath{S^{*}}}
\newcommand{\base}{\ensuremath{S_b}}
\newcommand{\accept}{\textsc{accept}}
\newcommand{\rejectstale}{\textsc{reject-stale}}
\newcommand{\rejectverify}{\textsc{reject-verify}}

\title{Authority at Commit Time:\\
Reject-and-Rerun Semantics for Governed Agentic Systems}
\author{Jesus Salas\thanks{The author is employed by Microsoft. This work was
performed independently, outside the scope of that employment, on personal time
and equipment. It does not represent the views or positions of Microsoft
Corporation.}\\
Independent Researcher\\
\texttt{jesus.salas@gmail.com}}
\date{August 2026}

\begin{document}

\maketitle

\input{sections/00_abstract}
\input{sections/01_introduction}
\input{sections/02_problem}
\input{sections/03_mechanism}
\input{sections/04_method}
\input{sections/05_results}
\input{sections/06_discussion}
\input{sections/07_limitations}
\input{sections/08_related_work}
\input{sections/09_conclusion}

\bibliographystyle{plainnat}
\bibliography{references}

\appendix
\input{sections/appendix_protocol}
\input{sections/appendix_claim_boundaries}

\end{document}

%% file: sections/00_abstract.tex
\begin{abstract}
Enterprise agents can compute for seconds or hours from a snapshot of policy,
facts, task state, models, tools, and verifiers that may change before their
work reaches the world. Completion alone therefore cannot confer institutional
authority. We treat agents as proposal producers and a logically authoritative
service as the sole authority for governed effects. At admission, the service
compares a proposal's declared dependencies and component manifest with current
state, applies the current verifier, and either reserves an idempotent effect
for outbox dispatch or rejects affected stale work for resynchronization and
rerun. Unaffected work may continue; lost responses resolve through canonical
verdicts and receipts.

The protocol composes optimistic concurrency control, semantic verification,
and transactional outbox dispatch. Its distinctive read set includes normative
authority and the model components that produced the proposal. We evaluate the
design through an implemented governed-work lifecycle, a local authority,
outbox, and reconciliation composition, and a multi-process fallback that
delegates effect authority when complete mediation is unavailable. The local
composition preserved selective fallback, unaffected work, retry and restart
recovery, and package reconstruction. Complete mediation produced 18 authority
commands and 18 external effects. Direct-credential arms produced eight
additional effects, all detected with independent journal reconciliation and
none falsely detected in the mediated control.

The results support commit-time reject-and-rerun semantics and expose the
clock, lease, ordering, and quorum obligations introduced by delegated
admission. They do not establish prevention or reversal of bypass
effects, external-journal completeness, production throughput, wide-area
availability, semantic completeness, or verifier correctness.
\end{abstract}

%% file: sections/01_introduction.tex
\section{Introduction}

An enterprise may run thousands of agents at once. Each agent can read policy,
facts, task state, models, and tools, then spend seconds or hours producing a
proposal. During that interval, any governing input may change. A policy may be
amended, a customer fact corrected, a task canceled, a verifier replaced, a
model withdrawn, or a prior effect recorded by another worker. The resulting
proposal may still look reasonable. It may even produce the same output that a
current worker would produce. The institutional question is not whether the
proposal looks plausible. It is whether the proposal is admissible now.

The simplest answer is to deny agents authority over effects. Agents may compute
concurrently and optimistically, but every proposal returns to a logically
authoritative service. That service owns the current ordered state and the only
commit path. It compares the proposal's declared base state with current state,
checks whether any changed dependency governs the task, applies the current
verifier, and either records an admitted effect for idempotent dispatch or
returns a structured refusal. Confirmed dispatch produces the completion
receipt. Affected stale work is rejected and rerun; it is not silently promoted
into current work.

This structure is optimistic concurrency control (OCC), not a new concurrency
algorithm. The agent's governing dependencies form a read set; proposal
submission begins validation; a stale dependency aborts the attempt; and rerun
is retry \citep{kung1981optimistic}. The extension is institutional. The read
set contains authority versions and component manifests for models, adapters,
prompts, tools, and verifiers, not only database objects. Admission also applies
a semantic verifier, and the requested write may be an irreversible external
effect rather than a rollback-safe database update. Complete mediation is
therefore a central precondition, not an implementation detail.

This placement matters. If a disconnected worker can continue to authorize
effects from cached state, the architecture needs leases, local cutover clocks,
fences, and rules for split epochs. If the worker can only submit proposals,
disconnection pauses institutional progress but does not create stale effects.
Availability then remains important, but it is an ordinary recovery-time and
recovery-point objective for the authoritative service rather than a second
source of truth at every agent node.

This paper extends the governed-work lifecycle of \citet{salas2026correct} and
the learned-artifact change controls of \citet{salas2026change}. The former
separates correctness from decision, execution, and change provenance. The
latter shows why recorded lineage cannot certify its own completeness and why
retained learned artifacts still require current-contract verification. Here we
move the same reasoning to prospective effects: a proposal has no institutional
standing until current authority admits it.

The contributions are:

\begin{enumerate}[leftmargin=*,itemsep=2pt]
  \item an explicit OCC formulation for institutional work that separates
  concurrent computation from authoritative effect;
  \item a reject-and-rerun protocol over versioned authority, facts, tasks,
  component manifests, evidence, verifier decisions, and completion receipts;
  \item partition, lost-acknowledgment, and scoped-federation semantics that do
  not delegate effect authority to agent nodes; and
  \item a reconciled evidence account combining an implemented Matrix
  governed-work ladder, a local networked authority/outbox/reconciliation
  composition, and a controlled delegated-gateway study whose stress failures
  clarify the value of the simpler boundary.
\end{enumerate}

The paper makes no claim that centralization removes distributed-systems work.
The authoritative service may be replicated, partitioned by governed scope, or
implemented over a consensus-backed database. The claim is logical: for each
effect scope there is one current ordered authority, and no agent-side state can
commit around it.

%% file: sections/02_problem.tex
\section{Problem and Invariant}

\subsection{Proposals are not effects}

Let an agent begin task $q$ from snapshot \base. Its output is a proposal
$p=(q,\base,D,M,E,a)$, where $D$ is the declared governing dependency set, $M$
is the model, adapter, tool, and verifier manifest used during work, $E$ is the
evidence bundle, and $a$ is the requested action. Let \current{} be the
authoritative state when the proposal reaches admission.

The agent is allowed to produce $p$ from an old snapshot. It is not allowed to
turn $a$ into an effect. Only the authoritative service can do that. This gives
the central invariant:

\begin{quote}
\textbf{Authoritative-commit invariant.} A proposal becomes an institutional
effect only if its governing dependencies are current at admission and the
current verifier admits it. Otherwise no effect is committed.
\end{quote}

An old global snapshot identifier is not automatically fatal. If state changed
only in an unrelated scope and every governing head in $D$ remains current,
normal verification may proceed. Conversely, a matching coarse epoch is not
sufficient if one declared governing head or required component hash differs.
The comparison is scope-sensitive and current-state authoritative.

\subsection{Why post hoc cleanup is insufficient}

A proposed message, payment, deployment, or entitlement change does not yet
exist as an effect, so retrospective revalidation cannot govern it. Admission
must happen before the authority-controlled adapter commits it. Once an
external effect has occurred through an ungoverned path, later ledger entries
can record, compensate, or escalate it, but cannot make the original effect
unhappen. The protocol therefore assumes complete mediation: every governed
effect path terminates at an adapter controlled by the authoritative service.

An external effect need not be a database write. Transmitting governed data
across a trust boundary, including through a model, retrieval request, or
nominally read-only tool call, is itself an effect when disclosure cannot be
undone after dispatch. The same commit boundary must therefore cover governed
communications as well as state-changing operations. This paper does not claim
a general information-flow-control mechanism.

This does not require the service to execute all computation. It may dispatch
an idempotent outbox command to a payment processor, messaging service, code
host, or internal tool. The requirement is that the authoritative transaction
records the unique effect identity and controls dispatch. A worker cannot use a
side channel to bypass admission.

\subsection{OCC inheritance and the governance delta}

Classical OCC separates a transaction into read, validation, and write phases,
then backs up and retries transactions whose validation conflicts
\citep{kung1981optimistic}. The present protocol inherits that mechanism
directly:

\begin{center}
\begin{tabular}{@{}lll@{}}
\toprule
OCC & Governed proposal & Result \\
\midrule
Read set & Governing dependency and component set & Snapshot-bound work \\
Validation & Compare with current heads & Current or stale \\
Write & Authority-controlled effect commit & Receipted effect \\
Abort and retry & \rejectstale{} and rerun & New attempt from current state \\
\bottomrule
\end{tabular}
\end{center}

The contribution is the composition of three dimensions for agent-produced
institutional work, not three independently novel transaction mechanisms. The
distinctive extension is the read set: it binds normative authority and the
model, adapter, prompt, tool, and verifier versions that produced the proposal.
The second dimension inherits semantic concurrency control, in which
application-level knowledge participates in admission beyond version or
serializability checks \citep{garciamolina1983semantic}; version agreement is
necessary but not sufficient because the current verifier may still return
\rejectverify{}. The third inherits scalable messaging practice for boundaries
where local state and an external effect cannot share one distributed
transaction: record the intent locally, relay it at least once, and make its
substantive impact idempotent \citep{helland2007life}. The transactional outbox
is the named practitioner pattern \citep{richardsonTransactionalOutbox}. The
protocol composes these mechanisms with complete mediation and canonical
governance receipts at the authoritative commit boundary.

\subsection{The reject-and-rerun rule}

When a changed dependency affects $q$, the authority does not promote the stale
proposal into current work. It returns \rejectstale{} with the current snapshot
and a machine-readable reason. The worker then:

\begin{enumerate}[leftmargin=*,itemsep=2pt]
  \item invalidates the stale attempt and any uncommitted descendants;
  \item installs current authority, fact, task, verifier, model, adapter, and
  tool state required by the task;
  \item restarts from an authoritative input checkpoint; and
  \item submits a new attempt with a new attempt identity but the same stable
  task and effect identities where appropriate.
\end{enumerate}

This rule sacrifices stale computation, not institutional safety. For long
tasks, wasted compute can be reduced by intermediate checkpoints, change
notifications, or shorter speculative stages. Those are optimizations. They do
not weaken commit-time admission.

\subsection{Safety and liveness are separate}

If the authoritative service is unavailable, affected nodes pause submission
and effect execution. Work may remain queued or in local computation, but it
cannot be committed. High availability, quorum design, and geographic
replication determine how long progress pauses. They do not change the safety
rule. Releases and canonical receipts require an effective recovery point of
zero; losing an acknowledged authoritative commit would destroy the meaning of
the ledger.

%% file: sections/03_mechanism.tex
\section{Authoritative Commit Protocol}

\subsection{State objects}

The protocol instantiates the governed-work sets introduced in
\citet{salas2026correct}:

\begin{itemize}[leftmargin=*,itemsep=2pt]
  \item \textbf{AuthoritySet}: current policy, permissions, revocations, and
  decision contracts;
  \item \textbf{FactSet}: versioned governing facts and supersession links;
  \item \textbf{TaskSet}: obligations, definitions of done, permitted tools,
  and stable task identities;
  \item \textbf{CompletionReceiptSet}: immutable attempt and effect evidence;
  and
  \item \textbf{SupervisorVerdict}: pass, fail, incomplete, rework, block, or
  escalation under the current completion contract.
\end{itemize}

A snapshot manifest binds the relevant heads of these sets to component hashes
for the model, adapters, prompts, tools, and verifiers. A proposal envelope
names its base manifest, dependency subset, stable task identity, unique attempt
identity, idempotent effect identity, requested action, and evidence hashes.

\subsection{Admission transaction}

For proposal $p$, the authoritative service performs the following ordered
operation:

\begin{enumerate}[leftmargin=*,itemsep=2pt]
  \item authenticate the submitter and check its franchise for task $q$;
  \item resolve the current heads for every declared governing dependency;
  \item compare the proposal manifest with the current required component
  manifest;
  \item if a changed head or component affects $q$, append a stale-attempt
  verdict and return \rejectstale{} with a resynchronization plan;
  \item otherwise run the current verifier against the proposal and evidence;
  \item on verifier failure, append \rejectverify{} and a rework, block, or
  escalation obligation; and
  \item on success, atomically reserve the effect identity and append the
  accepted verdict and outbox command; and
  \item after idempotent dispatch, append the external completion receipt or a
  failure and reconciliation obligation.
\end{enumerate}

The atomic unit is the comparison of current governing heads, verifier result,
effect-identity reservation, and verdict-plus-outbox append. Linearizable
storage is one implementation route for this unit
\citep{herlihy1990linearizability}. The protocol does not require a single
physical machine.

The admission transaction is the authority linearization point. Once the
effect identity and outbox intent are recorded, a later amendment does not
retroactively stale that admitted decision; cancellation, if the governing
contract permits it, is a new governed decision. The relay executes the
already-authorized intent, so the residual admission-to-landing window is
dispatcher latency rather than model-runtime duration.

\begin{table}[H]
\centering
\small
\caption{Admission outcomes. No refusal path commits the requested effect.}
\label{tab:outcomes}
\begin{tabularx}{\linewidth}{@{}lXX@{}}
\toprule
Outcome & Condition & Required next state \\
\midrule
\accept & Dependencies current; current verifier passes & Effect identity and
outbox command are reserved; completion is receipted after dispatch \\
\rejectstale & Governing state or required component changed & Attempt is
invalidated; worker synchronizes and reruns \\
\rejectverify & Current verifier fails & Evidence-preserving rework, block, or
escalation \\
Duplicate & Effect identity already decided & Return the canonical prior
verdict or current receipt; do not dispatch a second effect \\
Unavailable & Authority cannot decide & Pause; do not fall back to local
admission \\
\bottomrule
\end{tabularx}
\end{table}

\subsection{Lost acknowledgments and idempotency}

A connection may fail after the authority commits its admission decision but
before the worker receives the response. This is not an ambiguous
institutional state. The ledger already contains the verdict and outbox state.
The worker retransmits the same effect identity, and the authority returns the
canonical verdict or current completion receipt instead of dispatching twice.
The protocol therefore promises at-most-one committed effect per identity and
replayable knowledge of whether the effect was accepted, pending, completed,
or sent to reconciliation. It does not promise that arbitrary external systems
are naturally idempotent; the authority-owned adapter must supply that property
or reconcile it explicitly.

\subsection{Partition and resynchronization}

An agent node that cannot reach authority pauses governed submission and effect
execution. On reconnection it sends its last accepted snapshot hash and receipt
sequence. The authority returns the signed current manifest and missing
canonical receipts. If the governing subset is unchanged, queued proposals can
continue to current verification. If it changed, affected attempts are
invalidated, current components are installed, and the work is rerun. A global
manifest change always forces the handshake; dependency comparison determines
which tasks must restart.

\subsection{Unified authority with scoped ledgers}

Scale does not require every action to share one physical log. A deployment may
partition authority into customer, jurisdiction, product, or risk scopes. Each
scope still has one authoritative ordered history, possibly replicated for
availability. A cross-scope proposal declares the current heads for every scope
it governs, and the commit coordinator verifies all of them. Cached discovery
records may use time-to-live values; authorization records may not. This is a
logical unification claim, not a prescription for one database topology.

%% file: sections/04_method.tex
\section{Implementation and Evaluation Method}

The paper combines three evidence layers. They answer different questions and are
not pooled into a single verdict.

\subsection{Layer A: implemented governed-work lifecycle}

Matrix is a prototype central governance service with an append-only JSONL
event log, global sequence numbers, SQLite materialized state, typed operation
validation, versioned facts, explicit supersession, dependency-directed
invalidation, reconvergence tasks, outcome verification, and replay. Worker
models propose observations and decisions. Deterministic code owns declared
contracts and materialized lifecycle state. Higher-authority operations are
role-checked under prototype-local identities.

We use two already completed rungs from the frozen governed-work ladder. Both
are self-authored synthetic vendor-onboarding workflows evaluated on three
Phi-4 seeds. They are mechanism demonstrations, not rate estimates.

\paragraph{L4: versioned change recovery.}
The workflow contains independent legal-identity and tax-evidence branches. A
tax standard is superseded from v1 to v2 after initial execution. The governed
path must invalidate the old tax receipt and prior closure, preserve the legal
receipt, rerun only tax evidence under v2, and close from current evidence. A
separate proposal-admission challenge omits a required fact reference. Matrix
must reject it, issue an exact bounded rework obligation, admit only the
corrected proposal, and issue no work if correction remains incomplete.

\paragraph{L5: composed governed closure.}
The workflow composes legal identity, tax evidence, security review, and vendor
activation. The current proposal must be admitted before tasks issue; failed
attempts must remain in history; preconditions must hold; branch receipts must
be verified; the v1 to v2 tax change must preserve legal and security work;
and only a final activation receipt under current composite readiness may close
the workflow. The route-specific contract contains 20 lifecycle assertions per
seed.

The primary measurements are proposal-admission behavior, task execution before
admission, invalidated and preserved branch identities, rerun counts, current
receipt closure, replay, and per-action assertions. Both direct controls receive
the same authored amendment and target outcome, but they do not receive Matrix
admission, receipt, or selective-invalidation machinery.

\subsection{Layer B: local networked composition and reconciliation}

A separately frozen v2 study composes the paper's central mechanisms across
independent local HTTP processes backed by SQLite WAL. The authority service
authenticates proposal envelopes, resolves epoch- and scope-specific component
admission, runs pinned VAL, and atomically appends an admission verdict plus an
outbox command. A relay dispatches stable effect identities to an independent
external-effect service. A reconciler compares that service's journal with the
authority's outbox and completion records. The workload replays previously
recorded, VAL-admitted Qwen3-14B proposals; it is an integration workload, not a
new model-capability evaluation.

The three arms are fully mediated with reconciliation, direct-credential
bypass without reconciliation, and the same bypass schedule with
reconciliation. Frozen cases cover component revocation and fallback, an
unaffected specialist, verifier refusal, a lost acknowledgment and duplicate
retry, relay restart, eight observable direct bypasses, and six reconstruction
damage classes. Claim-bearing gates require no stale, verifier-rejected, or
duplicate mediated effects; current-work liveness; detection of all eight
journaled bypasses within 250 ms; zero false bypass detections in the mediated
arm; and refusal of every damaged package. The 250 ms bound is a local harness
gate, not a production service-level objective.

This layer tests a residual control, not an alternate commit path. A direct
bypass has already produced its effect. Reconciliation may detect the
discrepancy only if the external journal is complete for the governed scope; it
does not prevent, reverse, or compensate that effect.

\subsection{Layer C: fallback under incomplete central mediation}

Some deployments cannot enforce complete central mediation. Agents may require
locally scoped tool credentials, third-party systems may expose only local
gateways, or operation may need to continue while disconnected. When those
rights can be mediated locally but not centrally, effect authority is delegated
to cached edge state and the central protocol's safety argument no longer
applies directly. If a credential bypasses every admission point, neither layer
provides a safety guarantee.

A separate frozen study tested this fallback at three gateway processes backed
by a three-member local etcd control plane. The primary arm combined a versioned
release, local clocks, an impact-scoped temporal fence, release observation,
gateway leases, atomic effect identities, and verified fallback. A sealed oracle
clock scored timing without being visible to gateways.

This study characterizes the cost of delegation rather than serving only as a
warning case. The healthy run asks whether the bundle composes under
its assumptions. Two stress tests ask what happens when a local clock exceeds
its declared error bound and when the control plane loses a majority. A
separately frozen confirmation timestamps the actual effect inside an
oracle-blind executor. The evidence must retain five separate verdicts: healthy
bounded run, original clock probe, clock-effect confirmation, split-epoch
conformance, and majority-loss stress.

\subsection{Evidence boundary}

Layer A implements most governed lifecycle semantics but does not constitute a
production networked commit service. Layer B implements a local networked
proposal, authority, outbox, retry, restart, and reconciliation path, but it
does not establish production identity, WAN behavior, cross-ledger atomicity,
or external-journal completeness. Layer C uses real processes and etcd, but
evaluates the fallback when effect authority must exist outside complete
central mediation. The protocol claim is therefore an architectural
consequence supported by bounded component evidence and a boundary comparison,
not a claim that every production property has been experimentally established.

\begin{table}[H]
\centering
\small
\caption{What each evidence layer can establish.}
\label{tab:evidence-layers}
\begin{tabularx}{\linewidth}{@{}lXX@{}}
\toprule
Layer & Establishes & Does not establish \\
\midrule
Matrix L4/L5 & Admission before task issue; current receipt closure; selective
invalidation; rerun; preservation; replay & Network partitions, production
identity, federated commit, throughput \\
Local composition & Current component admission, outbox dispatch, retry and
restart, journaled-bypass detection, reconstruction & WAN behavior, production
identity, journal completeness, prevention or reversal of bypass \\
Delegated fallback & Multi-process bundle behavior under frozen healthy and
stress conditions & Correctness of the central protocol or WAN availability \\
Protocol analysis & Required state, outcomes, and safety boundary & Empirical
operating rates or semantic verifier quality \\
\bottomrule
\end{tabularx}
\end{table}

%% file: sections/05_results.tex
\section{Results}

\subsection{Central lifecycle mechanisms}

Table~\ref{tab:matrix-results} reports the canonical L4 and L5 results. In L4,
the deliberately incomplete initial proposal was rejected on all three seeds,
the exact missing reference was returned as rework, and the corrected proposal
was admitted. The persistent-failure regression admitted neither attempt,
issued no task, executed nothing, and escalated with the unresolved obligation.
After the tax authority changed, only the dependent tax task reran; the legal
task and receipt remained current.

In L5, all three live Phi proposals were complete, so admission correctly
avoided unnecessary decision rework. Each run preserved a failed tax-attempt
receipt, verified legal, tax, and security evidence, invalidated the v1 tax
receipt after supersession, preserved the legal and security receipts, reran tax
under v2, and issued activation only from current composite readiness. All
60 route-specific assertions passed. A regression that removed one preservation
assertion held the run despite a green final business state, showing that
outcome appearance alone could not satisfy the lifecycle contract.

\begin{table}[H]
\centering
\small
\caption{Implemented Matrix lifecycle results. Counts are deterministic
assertions over three self-authored Phi-4 seed runs.}
\label{tab:matrix-results}
\begin{tabularx}{\linewidth}{@{}lrrX@{}}
\toprule
Measurement & L4 & L5 & Reading \\
\midrule
Final proposal admitted & 3/3 & 3/3 & Execution followed successful admission \\
Current final state reached & 3/3 & 3/3 & Both governed and direct arms reached the authored outcome \\
Selective affected-branch rerun & 3/3 & 3/3 & Tax reran under v2 \\
Unaffected branches preserved per run & 1 & 2 & Legal; then legal plus security \\
Governed attempts per run & 4 & 6 & Direct control used 5 and 8 \\
Unnecessary reruns avoided per run & 1 & 2 & A mechanism count, not an enterprise rate \\
Route-specific assertions & n/a & 60/60 & 20 assertions on each of three routes \\
Replay and current-receipt closure & 3/3 & 3/3 & Stale receipt did not close the work \\
\bottomrule
\end{tabularx}
\end{table}

These results establish that the implemented prototype can treat an agent
output as a proposal, withhold work until admission, invalidate affected proof,
rerun only declared dependent work, and require a current receipt for closure.
They do not establish that the dependency graph is complete. As Paper 3 argues,
lineage is a candidate-scope mechanism; sound retention still depends on an
independent current contract \citep{salas2026change}.

\subsection{Local composition and independent reconciliation}

The v2 local networked study passed every frozen commit-safety, kernel-liveness,
reconciliation, and reconstruction gate. Table~\ref{tab:reconciliation-results}
separates what appeared in the authority outbox, what appeared in the external
journal, which effects used a direct credential, and which discrepancies the
reconciler reported.

\begin{table}[H]
\centering
\small
\caption{Local authority-commit and reconciliation results. Counts are
deterministic scenario outcomes, not enterprise rates.}
\label{tab:reconciliation-results}
\begin{tabularx}{\linewidth}{@{}Xrrrr@{}}
\toprule
Arm & Authority outbox & External effects & Direct effects & Detections \\
\midrule
Fully mediated + reconciliation & 18 & 18 & 0 & 0 \\
Direct bypass, no reconciliation & 18 & 26 & 8 & 0 \\
Direct bypass + reconciliation & 18 & 26 & 8 & 8 \\
\bottomrule
\end{tabularx}
\end{table}

In both bypass arms, the authority ledger remained internally reconstructable
while omitting all eight direct effects. In the claim-bearing reconciled arm,
the independent journal exposed every omission: detection latency ranged from
60.607 to 130.249 ms against the frozen 250 ms local gate. The fully mediated
control produced no false bypass detections.

The same study preserved the composed safety and liveness properties. All eight
amendment-affected tasks rejected the revoked specialist and committed once
through the current fallback; all eight unaffected tasks retained the admitted
specialist and committed once. A deliberately damaged plan produced no effect.
One lost acknowledgment per arm caused a retry without a duplicate external
journal entry, and a fresh relay drained the final pending command after
restart. Historical v1 receipts remained byte-identical. Every unmodified
package reconstructed from its sealed records, while mutation or deletion of
each of six authority, outbox, completion, external-journal, reconciliation, or
component-amendment record classes caused refusal.

This is controlled composition evidence. The proposals were recorded model
outputs, the services ran on one host, and the external journal was treated as
complete by construction. Detection neither prevented nor reversed a bypass.

\subsection{The measured cost of delegated admission}

The delegated-gateway bundle passed its healthy bounded-clock study. Across 16
scenario-arm runs it recorded 103 receipts and 60 commits, with zero stale,
premature, duplicate, verifier-bypassing, or prohibited-compensation effects.
All 15 paired boundary and scoped-admission probes passed. Partition recovery
was 44.074 ms, verified fallback added exactly two commits, and the observed
healthy split epoch was 174.251 ms under the frozen 385 ms maximum.

The stress evidence prevents a flattened success reading:

\begin{itemize}[leftmargin=*,itemsep=2pt]
  \item the original out-of-bound-clock probe was
  \texttt{PREDICTION\_NOT\_OBSERVED} because response timing did not establish
  the effect time;
  \item the separately frozen confirmation instrumented the actual effect and
  observed the time-only gateway commit at oracle 774.363 ms, 25.637 ms before
  $T=800$ ms;
  \item the matched release-observation bundle refused that proposal before it
  had observed v2, but this does not show event-based safety after observation
  and does not attribute the refusal to its live lease;
  \item a delayed last-gateway schedule produced a 340.597 ms split epoch inside
  the frozen $[315,385]$ ms conformance interval for the derived ceiling; and
  \item majority loss preserved safety but failed liveness, recovering in
  545.133 ms against the preregistered 500 ms deadline.
\end{itemize}

The result is not that etcd or leases are defective. When complete mediation is
unavailable, the tested bundle is a positive fallback: under its healthy
assumptions it preserved the declared safety properties while allowing scoped
progress. The cost is explicit. Delegating effect admission to cached gateway
state creates clock, observation, lease, and quorum behaviors, including the
observed stress failures. Under complete authoritative commit, an isolated node
has no local effect privilege to preserve. It pauses and later resynchronizes.

\subsection{The requested property and the remaining gap}

The work achieves the conceptual property sought: stale computation may exist,
but nothing becomes a governed effect without current authoritative
verification. The implementation evidence covers the core lifecycle around
admission, invalidation, rerun, receipts, closure, local networked outbox
dispatch, retry, restart, and journaled-bypass detection. It does not close the
full production claim. Authenticated production identities, signed manifests,
independently guaranteed journal coverage, WAN failover, throughput, and
cross-scope commit remain to be implemented and measured as a system.

%% file: sections/06_discussion.tex
\section{Discussion}

\subsection{The architecture is intentionally asymmetric}

Agents may be numerous, heterogeneous, stale, or temporarily disconnected. The
authority cannot. This is not a claim that one physical server must be perfect.
It is a claim that institutional state has a single logical ordering at each
governed scope. Replication may keep that ordering available. It must not create
multiple independent admission truths.

This asymmetry turns a difficult safety problem into a familiar availability
problem. If authority is unreachable, business progress pauses. For a bank or a
large enterprise, that is serious, but it is already handled as service
continuity: redundancy, failover, disaster recovery, capacity planning, and
regional routing. Allowing workers to commit from stale cached authority does
not solve continuity. It trades a visible outage for silent institutional
divergence.

\subsection{Complete mediation is the decision boundary}

The central design is preferred when the institution can ensure that every
effect passes through an authority-owned adapter. Where that precondition cannot
be enforced, the delegated mechanism is not merely an inferior architecture. It
is the fallback needed to govern local effect authority. The v1 evidence then
has a positive reading: fences, release observation, leases, effect identities,
and fail-closed quorum behavior can compose under declared assumptions. Its
stress results quantify why those assumptions and separate liveness reporting
remain necessary.

This yields a deployment rule. Centralize commit whenever complete mediation is
available. Delegate only the minimum effect scope that must operate without it,
and carry the measured fencing and availability obligations with that delegated
right. Where separately credentialed paths can still reach the target system,
reconcile against an independently controlled, complete external journal. That
control detects a declared class of omissions after the effect; it is not a
substitute for mediation when prevention matters.

The reconciliation result has the same information shape as the learned-
artifact result in Paper 3. A lineage graph cannot certify an edge it never
recorded, and an authority ledger cannot discover an effect that bypassed every
authority-owned record. Independent current-contract revalidation supplies the
first external check; an independently controlled target journal supplies the
second. Neither removes completeness assumptions. The latter relocates them to
journal coverage.

\subsection{RTO and RPO}

The acceptable outage duration is a recovery-time objective. It depends on the
business and effect class. The ledger's acknowledged releases and receipts have
a stricter recovery-point objective: effectively zero. A deployment can queue
work while admission is unavailable, prioritize high-risk scopes, or route to a
healthy replica. It cannot declare stale local state authoritative merely to
meet a latency target.

Long-running agents introduce an economic exposure window. More work can become
obsolete before submission. Intermediate authoritative checkpoints, dependency
subscriptions, and component-version notices can reduce this waste. The safety
question for that attempt is resolved at the authoritative admission
transaction. The remaining delay before an accepted effect lands is
outbox-dispatch latency, which a production deployment must bound and monitor,
rather than the full model runtime.

\subsection{Reuse is an optimization, not a commit path}

If a changed authority, fact, task, model, or verifier affects a proposal, the
proposal is no longer the work requested by the current institution. Rejection
and rerun from a current checkpoint is therefore the safe default. A domain may
preserve an unaffected intermediate artifact only through an explicit
current-authority admission rule. Reuse can reduce recomputation, but it is a
new governed operation and cannot make stale work current by assertion.

\subsection{Model updates are governing changes}

A model, adapter, prompt, tool, or verifier can be part of the task's governing
manifest. If a required component changes during a partition or long run, the
authority rejects the old-manifest proposal when that component governs the
task. The worker installs the current manifest and reruns. If the model is
unchanged and only an unrelated scope moved, the worker need not reload it.
This makes model synchronization explicit without treating the model as the
authority.

\subsection{Receipts resolve lost responses}

The case where a commit succeeds but the response disappears is often described
as an unavoidable race. At the institutional layer it is a receipt lookup. The
canonical ledger records whether effect identity $e$ was accepted. A retry with
$e$ returns that answer. The hard boundary moves to the adapter between the
authoritative outbox and the external system. Exactly-once business semantics
there require an idempotency key, a transactional integration, or explicit
reconciliation. In the local composition, one lost acknowledgment per arm was
retried without duplicating the external journal entry, and a replacement relay
drained a pending command. Those cases demonstrate the implemented identity and
recovery path under the harness, not natural exactly-once behavior in arbitrary
external systems. The property cannot be inferred from an agent response.

\subsection{Connection to the research arc}

Paper 1 defines what governed execution must prove
\citep{salas2026correct}. Paper 2 shows that admitted governance records can
also become supervision within machine-verifiable workflow contracts
\citep{salas2026records}. Paper 3 governs learned artifacts when authority
changes \citep{salas2026change}. This paper places those objects on one live
boundary: current authority decides whether a concurrent proposal becomes an
effect. Together, the arc moves from provenance, to learning, to artifact
change, to runtime commitment.

%% file: sections/07_limitations.tex
\section{Limitations and Open Work}

\subsection{Complete mediation and credential isolation}

Complete mediation is the strongest precondition in the paper. The current
protocol specifies that every governed effect terminates at an
authority-controlled adapter, but it does not itself enforce that topology. An
agent with direct credentials to an external system can bypass admission and
defeat the invariant without appearing in the ledger. Production deployment
therefore requires credential isolation, egress control, authority-owned tool
adapters, and independent audit for unmediated effects. The local composition
detected all eight deliberately journaled bypasses, but only after their effects
occurred. It assumes the external journal is complete for the governed scope;
unjournaled effects remain undetectable. Where central controls are impossible
but a local gateway can still mediate, the delegated fallback and its stronger
fencing assumptions apply. A credential that bypasses central admission, local
admission, and the declared journal remains outside every demonstrated claim.

\paragraph{Prototype evidence.}
The Matrix ladder is self-authored and uses three Phi-4 seeds. It demonstrates
implemented paths and regressions, not enterprise frequencies, human-review
cost, or production reliability. The delegated study uses real local processes
but not a wide area network. The v2 composition uses recorded Qwen proposals,
local HTTP services, and SQLite WAL on one host; its 250 ms detection deadline
is a harness gate rather than a service-level objective.

\paragraph{Protocol versus implementation.}
Signed snapshot manifests, authenticated production identity, independently
guaranteed journal coverage, federated commit, WAN behavior, production
throughput, and failover are not established by the current prototypes. Local
outbox idempotency, lost-acknowledgment retry, relay restart, and reconciliation
are measured only in the bounded v2 harness.

\paragraph{Contract correctness.}
Deterministic admission perfectly enforces only the declared contract. A bad or
incomplete contract can admit unsafe work or reject valid work. Independent
authorship, human adjudication, semantic-verifier validation, and monitored
appeals are required for open-world deployment.

\paragraph{Dependency completeness.}
The service can compare only governing dependencies it knows. An omitted edge
may leave a stale proposal apparently current. Current-contract verification
reduces this risk only for properties the verifier checks. This is the same
information boundary developed for learned artifacts in
\citet{salas2026change}.

\paragraph{Availability and scale.}
Logical authority may become a bottleneck or outage domain. Physical
replication and scoped ledgers are proposed, not benchmarked here. Cross-scope
transactions may require coordination and can sacrifice availability during
partitions. No throughput, tail-latency, regional-failover, or disaster-recovery
claim is made.

\paragraph{Compensation.}
Some external effects cannot be made atomic with the ledger. Compensation is a
new governed effect and can itself be prohibited by current authority. The
protocol therefore does not claim universal rollback.

The next claim-bearing experiment should move the networked admission envelope
to an independently administered environment and test authenticated identities,
guaranteed journal coverage, partition and failover behavior, throughput, and
cross-scope cases under frozen safety and liveness gates.

%% file: sections/08_related_work.tex
\section{Related Work}

\paragraph{Optimistic concurrency control.}
The protocol directly inherits OCC's read, validation, write, and abort-retry
structure \citep{kung1981optimistic}. Its contribution is not a new concurrency
algorithm. It extends the validated read set to normative authority and model
components, adds current semantic verification, and places potentially
irreversible external effects behind complete mediation.

\paragraph{Semantic concurrency control.}
Garcia-Molina showed how application-level semantic knowledge can admit
schedules beyond those allowed by purely syntactic conflict tests
\citep{garciamolina1983semantic}. The current verifier inherits that general
move: version validation and semantic admission are distinct checks. The
governance-specific object is a proposal whose read set also records the
authority and model-component provenance that produced it.

\paragraph{Beyond distributed transactions and transactional outbox.}
Helland treats an entity as a local scope of serializability and describes
transactional message intent, at-least-once delivery, and application-level
idempotency across scopes that cannot share one atomic transaction
\citep{helland2007life}. Richardson packages the local-state-plus-message form
as the transactional outbox pattern \citep{richardsonTransactionalOutbox}.
Here the outbox carries an admitted effect intent from the authoritative ledger
to an authority-controlled adapter. Idempotent effect identity and
reconciliation remain necessary because the external system is not part of the
ledger transaction.

\paragraph{Authoritative ordering and authorization snapshots.}
Linearizability formalizes the appearance of single-copy atomic operations in a
concurrent system \citep{herlihy1990linearizability}. Spanner demonstrates
globally distributed transactions and externally consistent ordering under
explicit time assumptions \citep{corbett2012spanner}. Zanzibar demonstrates a
globally distributed authorization system with consistent policy evaluation
\citep{pang2019zanzibar}. Its zookie encodes an authorization evaluation
snapshot so later reads and checks can request a causally appropriate snapshot.
The proposed manifest similarly carries governing versions, but also binds the
task's fact and component provenance and requires exact current-state validation
plus semantic admission before effect. This paper does not contribute a new
consensus or transaction algorithm.

\paragraph{Leases and coordination services.}
Leases bound cached rights in distributed systems \citep{gray1989leases}, and
Chubby provides coarse-grained locking and coordination for distributed clients
\citep{burrows2006chubby}. The earlier gateway study used the same family of
ideas through etcd. Under complete mediation, these mechanisms can remain
internal options for keeping the authority service available. Under delegated
admission, a gateway lease instead bounds a local right to admit, and the v1
study measures that more demanding role.

\paragraph{Long-running transactions and compensation.}
Sagas decompose long transactions into steps with compensating actions
\citep{garciamolina1987sagas}. Governed agent effects differ in one important
respect: a compensation is itself a proposed effect and current authority may
refuse it. The present protocol therefore treats compensation as a new
admission, not an automatic inverse.

\paragraph{Governed agent workflows.}
\citet{salas2026correct} distinguishes correctness from decision, execution,
and change provenance and defines the governed-work sets used here.
\citet{salas2026records} shows that verifier-admitted records can supervise
bounded workflow specialists. \citet{salas2026change} separates lineage-based
impact nomination from independent current-contract admission for learned
artifacts. The new contribution is prospective: a concurrent agent proposal
cannot acquire institutional standing until current authoritative commit.

%% file: sections/09_conclusion.tex
\section{Conclusion}

The safest place to resolve authority change is the place where a proposal
would become an effect. Agents may compute concurrently from snapshots, and
some of that computation may become stale. They remain non-authoritative. A
logically unified service compares the submitted checkpoint and dependencies
with current state, applies the current verifier, and alone can append the
admission verdict and outbox command. Confirmed dispatch produces the
completion receipt. Affected stale work is rejected, synchronized, and rerun.
It is not treated as current work by assumption.

Mechanically, this is OCC over institutional state. The novelty is not the
read-validate-write cycle. It is what becomes part of validation and what is at
stake at write time: current authority, facts, tasks, model components, semantic
verification, and a potentially irreversible external effect.

Matrix already implements the core lifecycle around this boundary: proposal
admission, versioned authority and facts, receipt-bound closure, selective
invalidation, rerun, preservation, supervisor verdicts, and replay. Its
composed synthetic rung passed all 60 declared lifecycle assertions across three
Phi-4 runs. A local networked composition then carried recorded model proposals
through component revocation, current verification, authority-owned outbox
dispatch, retry, restart, and independent reconciliation. Its authority ledger
could not discover eight direct-credential effects from internal records; an
independent external journal detected all eight in the reconciled arm, with no
false detections in the fully mediated control. The separate delegated-gateway
study shows what to do when complete central mediation is unavailable. Cached
local admission can be governed under healthy assumptions, but it carries
additional clock, lease, observation, and quorum obligations, including a
premature clock-driven effect and a majority-loss liveness failure under stress.

The resulting claim is deliberately bounded. We have a coherent protocol and
working lifecycle and local composition components, not a production federated
commit service. Reconciliation detects only effects represented in the declared
journal and cannot prevent or reverse them. The next systems result must measure
authenticated production identities, independently guaranteed journal coverage,
WAN failover, throughput, and scoped-ledger behavior. The architectural rule
should remain unchanged: nothing becomes governed merely because an agent
completed it; it becomes governed only when current authority accepts it.

%% file: sections/appendix_protocol.tex
\section{Protocol Field Sketch}
\label{app:protocol}

Table~\ref{tab:fields} lists the minimum claim-bearing fields. A production
encoding may add signatures, tenancy, tracing, retention, encryption, and
privacy controls.

\begin{table}[H]
\centering
\small
\caption{Minimum snapshot, proposal, and receipt fields.}
\label{tab:fields}
\begin{tabularx}{\linewidth}{@{}lX@{}}
\toprule
Object & Required content \\
\midrule
SnapshotManifest & Snapshot identity; ordered ledger heads; AuthoritySet,
FactSet, and TaskSet heads; model, adapter, prompt, tool, and verifier hashes;
issuer; issue time; signature \\
ProposalEnvelope & Stable task identity; attempt identity; idempotent effect
identity; base snapshot; declared governing heads; component manifest; requested
action; evidence and dependency hashes; worker identity \\
AdmissionVerdict & Accepted, stale, verifier-failed, duplicate, blocked, or
escalated; current snapshot; compared heads; verifier identity; reasons;
resynchronization or rework obligations \\
CompletionReceipt & Effect identity; attempt identity; task identity; admitted
action; current governing heads; verifier result; before and after state hashes;
outbox or external receipt; ledger sequence \\
ReconnectRequest & Last snapshot; last canonical receipt sequence; locally
queued task and attempt identities \\
ReconnectResponse & Current signed snapshot; missing canonical receipts;
invalidated attempts; tasks safe to continue; required component updates;
rerun checkpoints \\
\bottomrule
\end{tabularx}
\end{table}

\section{Reference Decision Procedure}

Given proposal $p$ and current state \current{}, the authority first resolves
any existing verdict for $p$'s effect identity. If one exists, it returns that
canonical verdict. Otherwise it authenticates the worker and reads every
declared governing head from \current{}. A changed head that affects the task,
or a mismatched required component, produces \rejectstale{}. If the proposal is
current, the authority evaluates the current verifier. Failure produces
\rejectverify{}. Success atomically reserves the effect identity and appends the
verdict and outbox command. An idempotent relay performs the external write and
then appends the completion receipt, or records a failure and reconciliation
obligation.

The procedure assumes that affectedness is defined by an admitted governance
contract, not guessed by the worker. Conservative deployment may reject on any
manifest change until finer dependency contracts have been independently
validated. Optimization changes wasted work and false blocks; it must not create
an alternate commit path.

\section{Reconnect Cases}

\begin{enumerate}[leftmargin=*,itemsep=2pt]
  \item \textbf{No relevant change.} Snapshot hash moved, but all governing
  heads for the queued task match. Continue to current verification.
  \item \textbf{Relevant authority or fact change.} Invalidate the attempt,
  install the current snapshot, and rerun.
  \item \textbf{Required model or tool change.} Install the new component,
  invalidate dependent work, and rerun.
  \item \textbf{Commit succeeded before disconnect.} Return the canonical
  receipt for the stable effect identity. Do not rerun the effect.
  \item \textbf{Verifier now fails.} Preserve evidence and issue rework, block,
  or escalation. Do not commit.
\end{enumerate}

%% file: sections/appendix_claim_boundaries.tex
\section{Claim-Control Ledger}
\label{app:claims}

\begin{table}[H]
\centering
\small
\caption{Permitted and prohibited readings.}
\begin{tabularx}{\linewidth}{@{}p{0.23\linewidth}XX@{}}
\toprule
Topic & Permitted & Not permitted \\
\midrule
Central invariant & No governed effect is admitted without current dependency
comparison and current verification, assuming complete mediation & Safety for
effects that bypass the authority \\
Matrix lifecycle & Implemented admission, invalidation, selective rerun,
receipt closure, verdict, and replay on self-authored fixtures & Production
networked commit, enterprise rates, or open-world completeness \\
Local composition & Implemented current component admission, authority-owned
outbox, retry, restart, reconstruction, and detection of all eight journaled
bypasses in the reconciled arm & Prevention, reversal, or compensation of
bypass; journal completeness; WAN or production latency \\
Delegated fallback & The tested bundle composed under healthy assumptions and
is a fallback where complete mediation is unavailable; stress exposed one
premature time-only effect and one liveness failure & Universal failure of
leases, etcd, or event-based activation \\
Reject and rerun & Affected stale proposals require new admission before reuse
& Optimal recomputation cost, automatic semantic affectedness, or implicit
reuse \\
Availability & Authority outage becomes fail-closed unavailability & Measured
WAN recovery, unlimited scale, or zero downtime \\
Idempotency & Stable effect identity makes ledger retry decidable & Natural
exactly-once behavior in every external system \\
Federation & Logical authority may be physically replicated or scoped & Proven
cross-ledger atomicity or a required single-server topology \\
Verification & Current verifier remains the acceptance authority & Verifier
correctness, policy validity, or objective truth \\
\bottomrule
\end{tabularx}
\end{table}

\paragraph{Evidence verdicts remain separate.}
L4 and L5 use the canonical Matrix ladder. The local composition uses the
separately frozen v2 protocol and passed its complete gate as
\texttt{COMPOSED\_RECONCILIATION\_GO}. Delegated evidence uses the frozen v1
and v1.1 protocols. The clock verdict remains
\texttt{PREDICTION\_NOT\_OBSERVED}, later confirmation stays separate, and
majority-loss liveness remains a failure. No verdict is transferred across
evidence layers.